\documentclass[12pt,preprint]{aastex}

\slugcomment{Not to appear in Nonlearned J., 45.}

\shorttitle{Ionisation  in atmospheres of very low-mass objects}
\shortauthors{Helling, Jardine, Witte, Diver}

\begin{document}


\title{Ionisation in atmospheres of Brown Dwarfs and extrasolar planets\\ 
       I The role of electron avalanche}


\author{Ch. Helling\altaffilmark{1,2} and  M. Jardine\altaffilmark{1}}
\affil{1: SUPA, School of Physics \& Astronomy, University of St Andrews, North Haugh, St Andrews,  KY16 9SS, UK}
\affil{2: KITP, University of California, Santa Barbara, CA 93106, USA}
\email{ch80@st-andrews.ac.uk}

\and

\author{S. Witte\altaffilmark{3}}
\affil{Hamburger Sternwarte, Gojenbergsweg 112, 21029 Hamburg,Germany}

\and

\author{D.A. Diver\altaffilmark{4}}
\affil{School of Physics and Astronomy, University of Glasgow, Glasgow G12 8QQ, UK }




\begin{abstract}
Brown Dwarf and extrasolar planet atmospheres form clouds which
strongly influence the local chemistry and physics. These clouds are
globally neutral obeying dust-gas charge equilibrium which is, on
short time scales, inconsistent with the observation of stochastic
ionisation events of the solar system planets.  We argue that a
significant volume of the clouds in Brown Dwarfs and extrasolar
planets is susceptible to local discharge events. These are electron
avalanches triggered by charged dust grains. Such intra-cloud
discharges occur on time scales shorter than the time needed to
neutralise the dust grains by collisional processes. An ensemble of
discharges is likely to produce enough free charges to suggest a
partial and stochastic coupling of the atmosphere to a large-scale
magnetic field.
\end{abstract}





\section{Introduction}

Brown Dwarfs and extrasolar planets have both low luminosity and high
surface gravity.  Their spectra indicate a complex chemistry due to
the low temperatures but high pressures in the atmospheres. This
chemical complexity has led to the understanding that condensation
processes produce clouds across a large range of effective
temperatures (T$_{\rm eff}$ [K]) ranging from late-M dwarfs, the L-
and T-dwarfs (Allard et al 2001, Tsuji 2005, Burrows et al. 2006,
Stephens et al. 2009) into the planetary regime (Ackerman \& Marley
2001, Fortney et al. 2008, Burningham et al. 2008). Irradiation and
star-planet interactions have recently added to the richness of the
local chemistry (e.g. Fossati et al. 2010, Spiegel et al. 2009)
imposing additional non-equilibrium effects (Vidotto et al. 2010 a,b).

This range of spectral types also shows emission at high energies that
cannot presently be explained. Quiescent X-ray emission at a constant
level of L$_{\rm X} \approx 10^{-3}$ L$_{\rm bol}$ and radio emission
at a constant level of L$_{\rm radio} \approx 10^{-7}$ L$_{\rm bol}$
are  present in solar mass stars and persists in stars of lower masses,
all the way down to around M7 (Berger et al 2010). Between M7 and M9
the quiescent level begins to decay, but sporadic, powerful flares are
still observed at even later spectral types. A similar slow drop in
chromospheric emission is also seen (Reiners and Basri 2002).

In stars like the Sun, this emission is reasonably well
understood. These stars generate a magnetic field within their
convective interiors. This field rises buoyantly through the interior,
to emerge at the stellar surface and extend into the outer
atmosphere. The convective motions generate waves along these field
lines that transmit some fraction of the convective energy into the
outer atmosphere where it is released in reconnection events. The
largest of these are observed as X-ray\footnote{The X-ray luminosity
results from thermal gyrosynchrotron radiation resulting from
thermally produced electrons only.}  and radio flares, while the smallest
serve to provide the quiescent level of X-ray emission required to
power the corona (Parker 1988). This process, however, relies on a
coupling between the convective fluid motions and the magnetic
field. Without this coupling, energy cannot be stored in the magnetic
field and hence cannot be released in the outer atmosphere to generate
coronal X-ray emission. As demonstrated by Mohanty et al (2002),
however, stellar objects between M5 and L6 are too cool for this
coupling to be effective. Thermal ionisation processes are simply not
sufficient to provide a high enough ionisation fraction. If these non-binary low
mass objects are so cool, then, how do we explain the observed X-ray
emission? The presence of this emission indicates an ionised
atmosphere, but the cause of this ionisation is unknown.

The key to this puzzle may lie in the dust clouds that
begin to form in low mass stars at around the same effective
temperature range where thermal ionisation processes die away. Within the
solar system, cloud-covered planets are observed to produce various
kinds of discharge events (e.g.~Yair 2008). On Earth, dust storms and
volcanoes are examples for strong intra-cloud discharging of dusty
gases. We argue in this paper that such discharge events are to be expected
in the dusty gases of extrasolar atmospheres, in analogy to their
solar system counterparts.  If they could raise the ionisation fraction
by a sufficient amount, over a large enough volume, then they may be able
to explain the presence of sustained quiescent X-ray emission in objects
as cool as M9 and the presence of large flares in even lower mass objects.

There is, however, an immediate problem with this suggestion, which is that
classical charge equilibrium arguments point to a rapid neutralisation of dust particles.
This would imply that these cool, dusty atmospheres are neutral, at least
on long time scales. Observations of solar system planets and
laboratory discharge events, however, suggest that on {\it short} timescales,
discharge events should be expected. We propose that some rapid process,
such as the electron avalanche believed to act in lightning bolts, may operate.
Assuming that dust grains can be charged for long enough for avalanches to occur,
they may be capable of providing local and intermittent ionisation. Depending on the
volume of the atmosphere that is susceptible to this process, and the frequency with which
it occurs, this may be capable of explaining a low level of magnetic field coupling
and hence quiescent X-ray emission, or only intermittent coupling, leading to occasional flaring.

In Sect~\ref{sec:cloudheight}, we first briefly summarise cloud
properties of Brown Dwarfs and extrasolar planets and we investigate
the geometrical extension of clouds depending on the fundamental
parameters T$_{\rm eff}$ and log(g). By assuming that the dust grains,
which compose the clouds, are charged, we discuss their effect on the
local charge balance in Sect.~\ref{sec:chbal}. In
Sect.~\ref{sec:where}, we demonstrate that dust-initiated electron
avalanches are a strong candidate to temporally increase the
local degree of gas phase ionisation in the atmospheres of Brown
Dwarfs and extrasolar planets.

\section{Dust in Brown Dwarf and exoplanetary atmospheres}
\label{sec:cloudheight}

We analyse results from the dust cloud model developed by Woitke \&
Helling (2003, 2004), Helling \& Woitke (2006), and Helling, Woitke \&
Thi (2008) describing consistently the formation of a stationary cloud
by homogeneous nucleation and dirty growth/ evaporation, including
gravitational settling, element depletion, and convective element
replenishment. In the following, we utilise {\sc Drift-Phoenix}
atmosphere models (Dehn 2007, Helling et al. 2008a,b; Witte, Helling
\& Hauschildt 2009) in which our detailed kinetic model of dust cloud
formation is coupled with a radiative transfer code (Hauschildt \&
Baron 1999, Baron et al. 2003).  This allows for a consistent solution
of the model atmosphere problem and the dust cloud model from which we
derive the geometrical extension of the cloud. Based on {\sc
Drift-Phoenix} model atmospheres, we have shown that clouds form in
atmospheres of Brown Dwarfs and extrasolar planets, and that they are
composed of chemically heterogeneous grains of various sizes. The
local grain size distribution is dominated by small grains at the
cloud top and by large grains at the cloud base. These dust grains
gravitationally settle and change their chemical composition and their
sizes during this journey through the atmospheric temperature gradient
until they evaporate.  The cloud's material composition, however,
is similar amongst the Brown Dwarfs and giant gas planets
(Helling, Woitke \& Thi 2008) except if the element abundances are
very low ([M/H]$<-4.0$; Witte, Helling \& Hauschildt 2009). The
geometrical cloud extension can change considerably with log(g).

\begin{figure}
\begin{center}
\includegraphics[angle=90,width=0.85\textwidth]{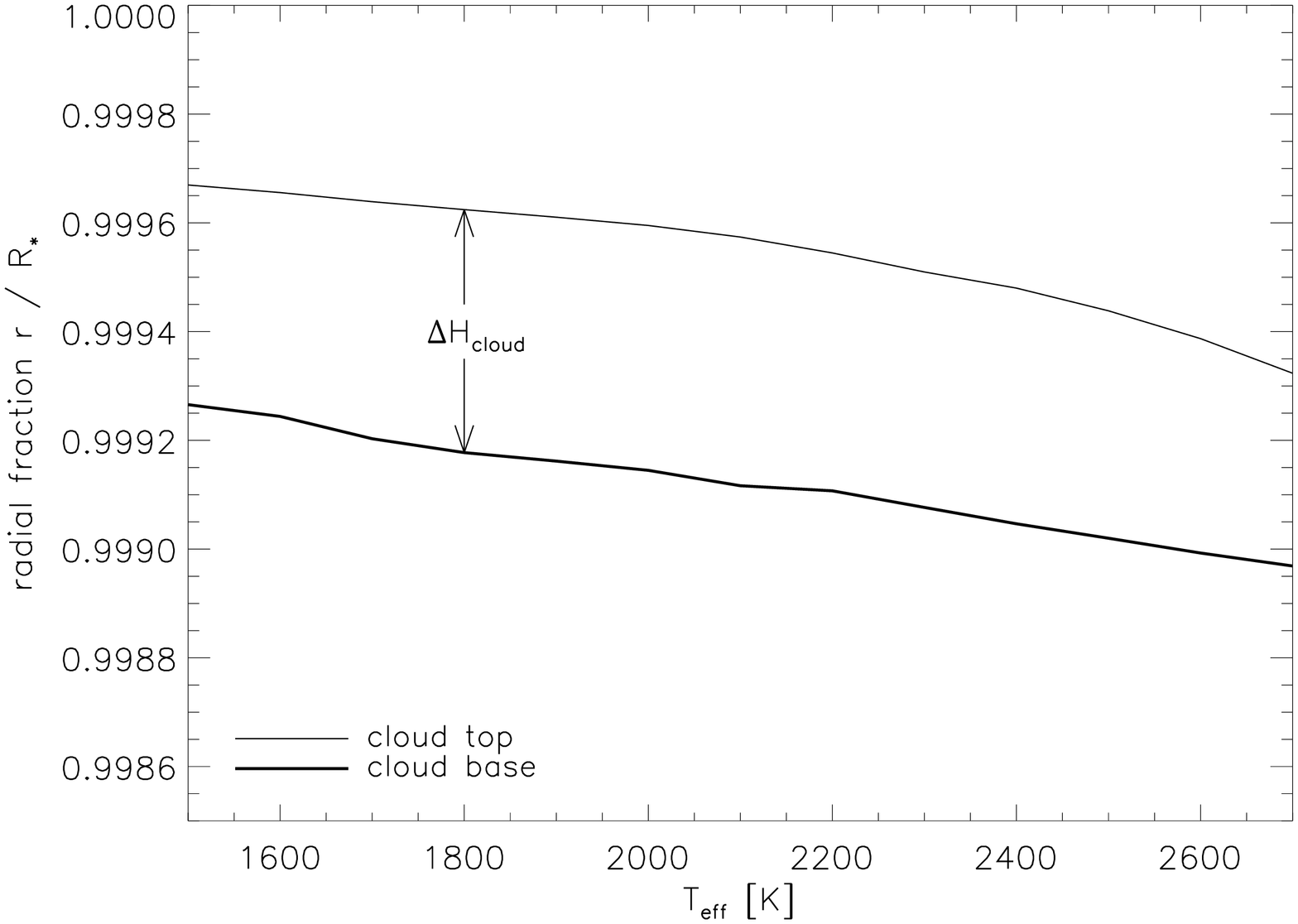}
\includegraphics[angle=90,width=0.85\textwidth]{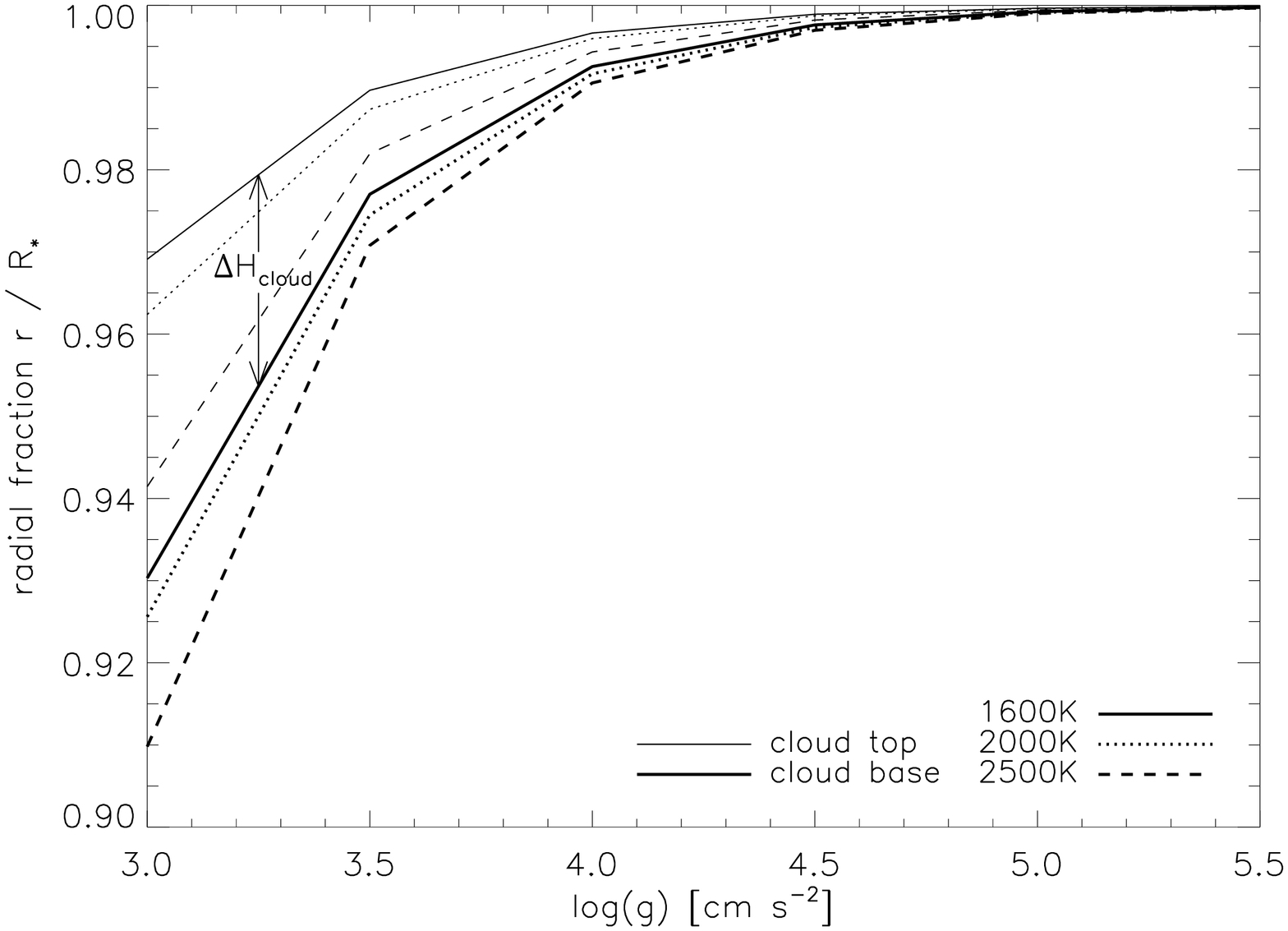}\\
\caption{\small Cloud extension $\Delta H_{\rm cloud} = (r_{\rm top} - r_{\rm base})\cdot R_*$ as function of T$_{\rm eff}$
(top, log(g)=5.0) and $\log$(g) (bottom) from {\sc Drift-Phoenix}
model atmospheres.  The cloud's top (@$r_{\rm top}$ [R$_*$], thin lines) is defined at the
fist nucleation maximum, the cloud base (@$r_{\rm base}$ [R$_*$], thick lines) is defined at
the location where all dust is evaporated (Fig.~2 in
Witte, Helling \& Hauschildt 2009).}
\label{fig:cloudthickness}
\end{center}
\end{figure}

The geometrical extension of the cloud is determined by the maximum
distance over which grains can fall before they evaporate.  The cloud
extension\footnote{ This definition is different to Woitke \& Helling
(2004). The geometrical thickness of the cloud layer was defined in
Woitke \& Helling (2004) as the degree of condensation of Ti $f_{\rm
cond}^{\rm Ti} = \epsilon^{\rm dust}_{\rm Ti}/\epsilon^0_{\rm Ti} >
1/e$ having decreased to $1/e$ at the cloud deck and the cloud base.
$\epsilon^{\rm dust}$ is the element abundance contained in the solid
phase and $\epsilon^0_{\rm Ti}$ the undepleted element abundance.} is
taken as the geometrical distance between the first nucleation maximum
(cloud top) and the point at which the dust has evaporated at high
temperatures (cloud base), hence $\Delta H_{\rm cloud} = (r_{\rm
top} - r_{\rm base})\cdot R_*$  where fractional distances $r$
are scaled to the radius $R_*$.  Figure~\ref{fig:cloudthickness}
(top) shows that the dependence of the cloud extension on the Brown
Dwarf's T$_{\rm eff}$ is comparably weak with a cloud extension of
$H_{\rm cloud}\approx 0.00036 \ldots 0.00041 R_*$. As an example,
a Jupiter-sized Brown Dwarf ($R_*=R_{\rm J}=7.1492\cdot\,10^9$cm;
equatorial Jupiter radius) with a surface gravity of $\log$(g)=5.0
would form clouds that spanned 27.5km i.e. 0.000385$R_*$
(Fig.~\ref{fig:cloudthickness}, top). Measuring cloud extension from
observations is, however, not straightforward as observations only
access the (frequency-dependent) optically thin part of the cloud
while the geometrical extension of the cloud layer is often much
larger (e.g., Banfield et al. 1998, Matcheva et al. 2005 for Jovian
clouds, Pont et al 2008 and Sing et al. 2009 for extra-solar clouds).

 The extension of the cloud changes much more significantly, however, with $\log$(g)
(Fig.~\ref{fig:cloudthickness}, bottom). It increases considerably
for young Brown Dwarfs and gas giant planets as they have a much lower
surface gravity of $\log$(g)$\approx 2.5 -4.0$ compared to old
Brown Dwarfs with $\log$(g)$>4.0$. This reflects the increasing
pressure scale height with decreasing gravity as $\Delta H_{\rm p}\sim
1/g $.  The cloud extension increases from $\Delta H_{\rm
cloud}=0.004\,\ldots0.005 R_*$ for $\log$(g)=4.0 to $\Delta H_{\rm
cloud}=0.03\,\ldots0.04 R_*$ for $\log$(g)=3.0 by almost a factor 10
(Fig.~\ref{fig:cloudthickness}). Table~\ref{tab:examples} shows
examples for Brown Dwarfs and WASP planets by assuming
that $\Delta H_{\rm cloud}$ varies only slowly with T$_{\rm
eff}$. Depending on the object's radius and surface gravity, the cloud
extension varies between some 10km and 5000km.

\section{Charge balance in dusty atmospheres}\label{sec:chbal}

 We therefore expect that Brown Dwarfs and extrasolar planets will
 form dust clouds whose geometrical extent depends mainly on the
 surface gravity. Although the presence of these dust clouds
 undoubtedly influences the local chemistry, it remains an open
 question whether they can significantly influence the large-scale
 dynamics. The observation of X-ray flares and coherent radio emission
 from Brown Dwarfs (Berger et al. 2010) suggests that they
 must possess a magnetic field. The presence of a global magnetic
 field opens up the possibility of coupling between this field and the
 gas in which it is embedded. In this case, by analogy with the Sun
 and solar-type stars, we would expect that the magnetic field may
 become twisted and tangled by the convective motions. This process
 stores magnetic energy throughout the field and it is the release of
 this energy by many reconnection events that is believed to power the
 persistent X-ray emission from the coronae of solar-like stars. 
 If, as on the Sun, Brown Dwarf flares are magnetically-powered,
 then there must have been at least a short-term, coupling between the
 magnetic field and the plasma.
 
The level of this field coupling is typically measured by the value of
 the magnetic Reynolds number $R_{\rm m}$ which is the ratio of the
 advective and diffusive contributions to the temporal evolution of
 the magnetic field. A magnetic field may be considered to be
 effectively coupled to the plasma if diffusion is negligible compared
 to advection and hence $R_{\rm m} > 1$. This requires, however, that
 there is a significant level of ionisation of the plasma, since the
 magnetic Reynolds number is proportional to the atmospheric
 ionisation fraction $f_{\rm e}=p_{\rm e}^{\rm tot}/p_{\rm gas}$, such
 that
\begin{eqnarray}
\label{eq:Reynr}
R_{\rm m} & = & u\,l\,\eta_{\rm d}^{-1}\\
          & = & u\,l \cdot \frac{4\pi\,q^2}{ m_{\rm e} c^2 }\frac{1}{\langle\sigma v\rangle_{\rm en}}\cdot f_{\rm e},
\end{eqnarray}
where $\eta_{\rm d}$ is the magnetic diffusivity, $q$ is the electric
 charge, $c$ is the speed of light, $m_{\rm e}$ is the mass of the
 free electron, and $\langle\sigma v\rangle_{\rm en}\approx 10^{-9}$
 cm$^3$s$^{-1}$ is the collisional cross section (velocity averaged
 momentum transfer cross section; Pinto \& Galli 2008). We
 choose $u=10^4$ cm/s as representative, large scale velocity value
 (Fig.~9 in Freytag et al. 2010), and $l=10^5$cm.
  
\begin{figure}
\begin{center}
\resizebox{16.5cm}{!}{\includegraphics{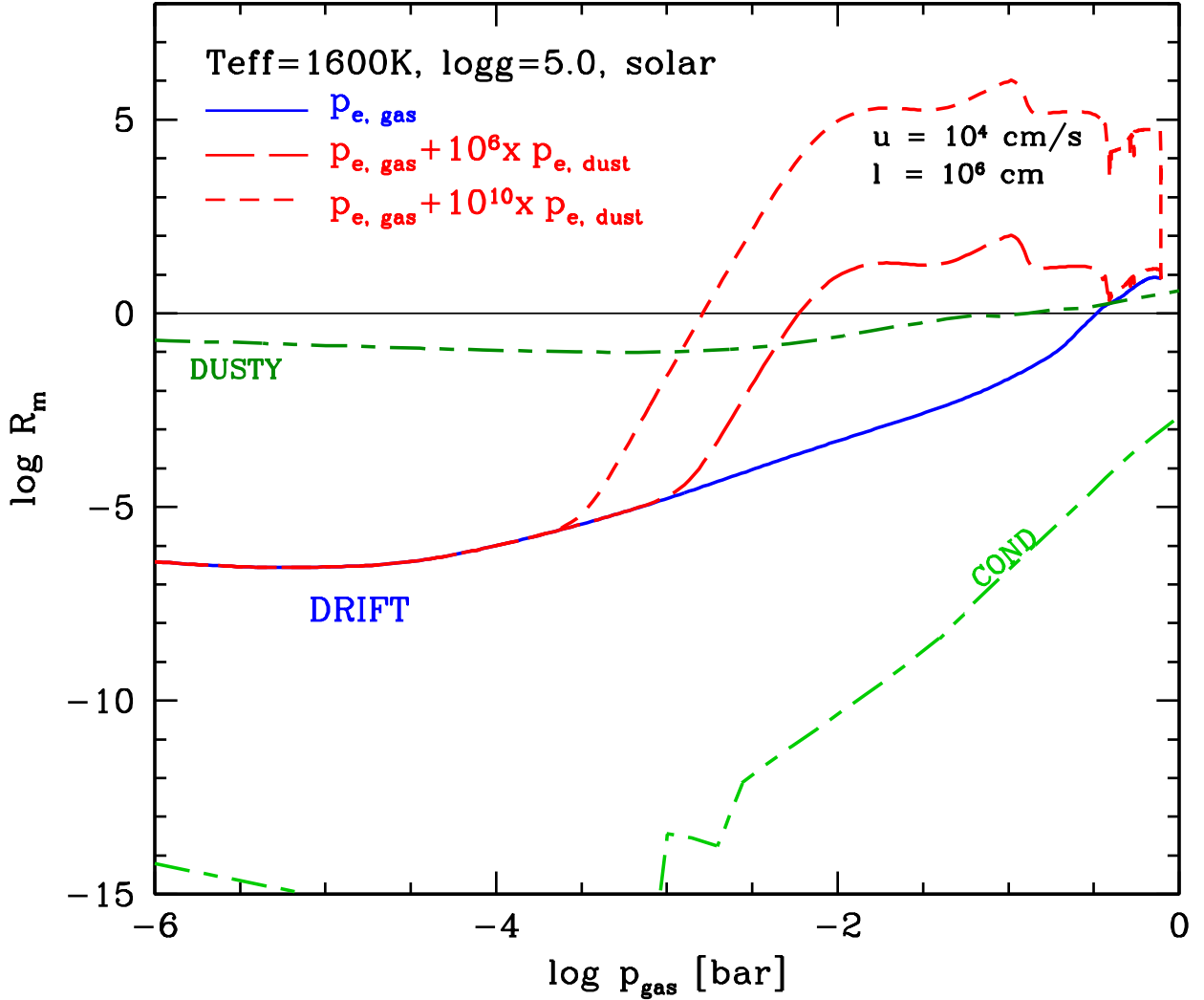}}
 \caption{\small Magnetic Reynolds number, $R_{\rm m}=u\,l\,\eta_{\rm
 d}^{-1}$, for a {\sc Drift-Phoenix} model atmosphere (blue) and a
 comparison with {\sc Dusty-Phoenix} (dark green) and {\sc
 Cond-Phoenix} (light green) results for thermal gas ionisation
 only. Magnetic coupling of the gas to a magnetic field is expected if
 $R_{\rm m}> 1$, hence above the black horizontal $R_{\rm m}=1$  line. Overplotted are the the cases with a
 charge production that has been artificially increased by a factor of $10^6$
 (long dashed line) and $10^{10}$ (short dashed line) inside the cloud. This
 demonstrates that magnetic coupling can be expected in limited parts
 of the dusty atmosphere if such a high number of charges can be produced.}
\label{fig:Rm_press}
\end{center}
\end{figure}

\begin{figure}
\begin{center}
\resizebox{16.5cm}{!}{\includegraphics{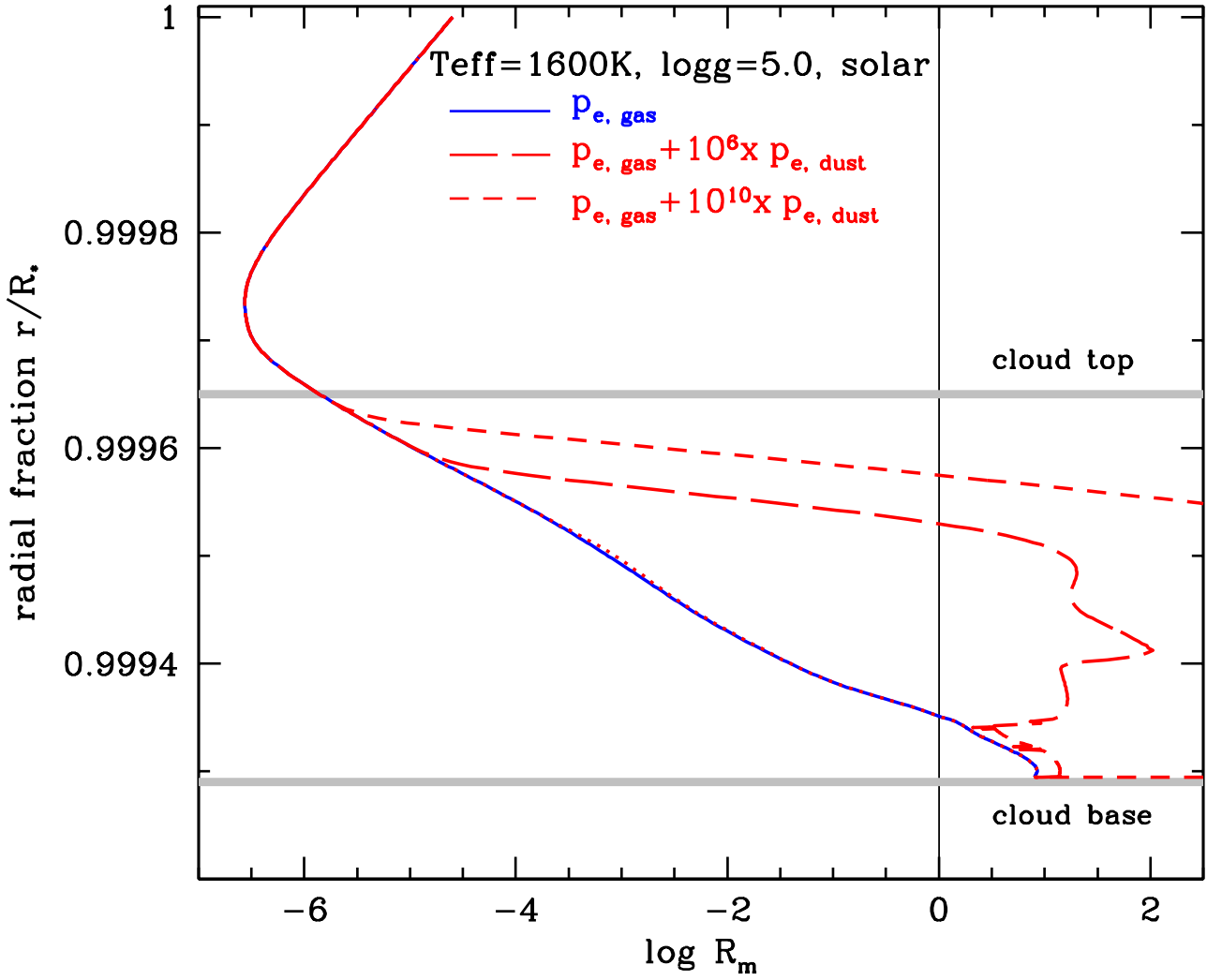}}
\caption{\small Magnetic Reynolds number, $R_{\rm m}=u\,l\,\eta_{\rm d}^{-1}$ in relation to the cloud height for the {\sc Drift-Phoenix} model atmosphere only. `Cloud top' and `cloud base' are defined as in Fig.~\ref{fig:cloudthickness}.}
 \label{fig:Rm_radfrac}
\end{center}
\end{figure}

 The magnetic Reynolds number and hence the degree of
 atmospheric coupling will therefore vary with depth
 in the atmosphere as the ionisation fraction varies. 
 Figure~\ref{fig:Rm_press} demonstrates this variation using the
 local gas pressure as a measure of atmospheric depth, while 
  Figure~\ref{fig:Rm_radfrac} describes its variation with
the cloud's radial fraction [r/R$_*$].  Figure~\ref{fig:Rm_press}
contains results of several different atmospheric codes.
The {\sc Drift-Phoenix}
results are compared to results for the {\sc Dusty-Phoenix} (green
dashed line, Fig.~\ref{fig:Rm_press}) and the {\sc Cond-Phoenix}
(brown dashed lines, Fig.~\ref{fig:Rm_press}) model
atmospheres\footnote{{\sc Dusty-Phoenix} and {\sc Cond-Phoenix} are
older versions of the {\sc Phoenix} mode atmosphere results where the
existence of dust was assumed rather than its formation
calculated. }. {\sc Dusty-Phoenix} produces a considerably higher
$R_{\rm m}$ due to thermal gas ionisation owing to an overall hotter
temperature in the atmosphere compared to {\sc Cond-Phoenix} and {\sc
Drift-Phoenix}\footnote{The different definition for the magnetic diffusivity
used by Gelino et al (2002) results in smaller magnetic Reynolds
numbers (black lines).}.

Only the very inner parts of the atmosphere reach $R_{\rm m}>1$ by
thermal gas ionisation in the {\sc Dusty-} and {\sc Drift-Phoenix}
model atmospheres. Thermal gas ionisation is, hence, not sufficient to
couple the gas inside and above the dust cloud to the magnetic field
as already shown by Gelino et al. (2002) and Mohanty et al. (2002). 
A significant increase in the electron number density would be
required to ensure that $R_{\rm m}>1$. In order to illustrate this
point, we show in Fig.~\ref{fig:Rm_press} that 
an increase of the electron number density by a factor $10^6$ per
cloud altitude would  be needed to ensure $R_{\rm m}>1$ (long dashed red line,
Fig.~\ref{fig:Rm_press}). This translates into $>$ 50\% of the cloud
(long dashed red line in Fig.~\ref{fig:Rm_radfrac}) being coupled to
any magnetic field present. This would be the inner $\sim 13$km of a cloud of
a Jupiter-sized Brown Dwarf (Sect.~\ref{sec:cloudheight}).  Note that
the atmospheric volume concerned increases with decreasing surface
gravity of the object, and hence, for young Brown Dwarfs or giant gas
planets (Sect.~\ref{sec:cloudheight}).

It is clear, therefore, that in order to maintain a state whereby most of the cloud is
ionised sufficiently to ensure an effective coupling to any magnetic field
present, we require a significant and sustained enhancement of the
electron number density. The magnitude of this enhancement can be 
illustrated quite simply by considering charge balance in the cloud.
 The electron density, $n_{\rm
 e}$, changes due to electron production processes ($I = I_{\rm gas} +
 I_{\rm dust}$) and electron recombination processes ($R= R_{\rm gas}
 + R_{\rm dust}$), such that, in equilibrium,
\begin{equation}
\label{eq:chcons}
\frac{d n_{\rm e}}{d t} =  I_{\rm gas} + I_{\rm dust} - R_{\rm gas} - R_{\rm dust}  = 0.\\
\label{eq:chequ}
\end{equation}
We neglect the thermal gas-phase processes, $I_{\rm gas}$ and $R_{\rm
gas}$, as we (also Gelino et al. 2002, Mohanty et al. 2002) have
demonstrated that thermal ionisation of the gas phase is negligible in
Brown Dwarf atmospheres.  If the electron production and electron
neutralisation are due to collisional processes with the dust grains
only, then
\begin{eqnarray}
I_{\rm dust}  &\sim& N n_{\rm d}^2 \pi a^2 v_{\rm rel},\\
R_{\rm dust} &\sim& n_{\rm e} n_{\rm d} \pi a^2 c_{\rm e},
\end{eqnarray}
with $n_{\rm d}$ the dust number density, $N$ the number of charges produced by a collisions, $a$ the dust grain radius, $v_{\rm rel}$ a relative velocity, and $c_{\rm e}$ the electron thermal velocity. Thus we simply have
\begin{equation}
\label{eq:IdRd}
\frac{I_{\rm dust}}{R_{\rm dust}} \sim N \frac{n_{\rm d}}{n_{\rm e}}\frac{v_{\rm rel}}{c_{\rm e}}.
\end{equation} 
The presence of dust can only significantly increase the electron
density if $I_{\rm dust}\gg R_{\rm dust}$ and, hence $N n_{\rm d}
\gg n_{\rm e}$.  Since the number density of dust grains is so much
less than that of electrons, we require that the number of charges $N$
produced by collisions must increase dramatically.  We estimate
that $I_{\rm dust}/R_{\rm dust} \sim 7\cdot 10^{-13} N$, hence we
require $N> 1.4\cdot 10^{12}$ in the low-pressure part at $p_{\rm
gas}\sim 10^{-5}$bar of the cloud\footnote{We have adopted values for
the low-pressure part of a cloud at $p_{\rm gas}\sim 10^{-5}$bar where
$T\sim 1000$K, $n_{\rm d}\sim 10^{-10}$cm$^{-3}$ (Helling et
al. 2008b), $v^{\rm drift}_{\rm rel}\sim 10^4$ cm s$^{-1}$ (Witte,
Helling, Hauschildt 2009), $n_{\rm e} \sim 0.076$ cm$^{-3}$ for
$p_{\rm e}\sim 10^{-14}$ dyn\,cm$^{-2}$. The electron thermal velocity
is $c_{\rm e}=\sqrt{2\,k\,T/m_{\rm e}}\sim 1.7\cdot 10^7$
cm\,s$^{-1}$.}. Conditions would
be more favourable at higher pressures as fewer free charges would be
needed: At a pressure of $10^{-2}$bar, $I_{\rm dust}/R_{\rm dust}
\sim 8.5\cdot 10^{-11} N$, and hence only $N> 1.2\cdot 10^{10}$ is
required\footnote{We have adopted values for
a part of the cloud with a higher pressure of $p_{\rm gas}\sim  10^{-2}$ bar
where
$T\sim 1150$K, $n_{\rm d}\sim 10$cm$^{-3}$ (Helling et
al. 2008b), $v^{\rm drift}_{\rm rel}\sim 10^2$ cm s$^{-1}$ (Witte,
Helling, Hauschildt 2009),  $n_{\rm e} \sim 6.3\,10^{5}$ cm$^{-3}$ for
$p_{\rm e}\sim 10^{-8}$ dyn\,cm$^{-2}$. The electron thermal velocity
is $c_{\rm e}=\sqrt{2\,k\,T/m_{\rm e}}\sim 1.9\cdot 10^7$
cm\,s$^{-1}$.}. These numbers are, however, sufficiently large that it
is unlikely that the simple collisional processes we have considered
will be sufficient to produce an equilibrium state with $R_{\rm m} >
1$.  We therefore expect that even in the presence of dust
clouds, the atmospheres of Brown Dwarfs should be globally
neutral. This does not, however, preclude short-term departures from
equilibrium which might (as in the solar system) lead to transient
ionisation events.

Concluding this section, we have demonstrated that a considerable
number of additional charges is needed to allow $R_{\rm m}>1$, or
equivalently, to provide a minimum degree of ionisation of $f_{\rm
e}>10^{-7}$. These charges need to be produced (or separated) on a
time scale shorter than a classical collisional recombination time
scale of grains in order to keep the atmosphere globally neutral over
a long time scale. 

\section{Electron avalanche in clouds}\label{sec:where}
 
We have argued that the consideration of thermal and dust-collisional
processes alone suggests a globally neutral atmosphere, but that
processes acting on shorter time scales can cause a violation of
strict charge equilibrium locally. This would be consistent with
intermittent  X-ray flares on Brown Dwarfs, where an rapid
release of energy produces intense X-ray emission for a short time,
followed by long periods of quiescence. We suggest electron avalanche
initiated by charged dust grains as one candidate for the fast
production of large amounts of free charges, and hence as source for a
locally increased degree of gas-ionisation in the atmospheres of Brown
Dwarfs and extrasolar planets.

We assume that dust grains are charged, as discharge events are
observable in a variety of environments (terrestrial thunder clouds,
dust devils, volcano plumes) and are expected to occur in others
(protoplanetary disks, Desch \& Cuzzi 2000).  These dust grains form
and have relative velocities e.g. by gravitationally settling and
turbulence which determines the cloud's geometrical extension in an
atmosphere (Woitke \& Helling 2003; Helling et al. 2006).
 
 Experiments have shown that a 100$\mu$m particle can carry $10^5$
 elementary charges (Sickafoose et al 2000). If we consider grains
 that are 3 orders of magnitude smaller, their charge load will be
 smaller by about the same order of magnitude. Helling, Woitke \& Thi
 (2008) have shown that the cloud's grain size distribution per
 altitude can be very broad in particular in low-gravity objects
 (log(g)=3.0) containing of order $10^4$ particles as large as
 10$\mu$m.  If these large particles would only release 1/100 of their
 accumulate $10^4$ charges into the gas phase before electron
 recombination, $>$ 50\% of the cloud would achieve $R_{\rm m}>1$ (see
 Fig.~\ref{fig:Rm_radfrac}). These charges may, however,
 neutralise by collisional recombination unless an avalanche
 instability like for instance in streamers can develop on a shorter timescale
 (see Sect.~\ref{ss:neutralis}).
 
 If enough charges remain on the grain surface (or none are released
into the gas phase), and the grains move with a relative velocity, a
strong electric field can be established that initiates a small-scale
discharge between the grains. A current forms that tries to establish
charge balance. Free thermal electrons gain enough energy when moving
through the grain's small-scale electric field to be able to ionise
the gas between the grains. Eventually the field disappears as the
grains move further apart. What remains is a partly ionised ambient
atmosphere while the charged grains escape due to gravitational
settling. We can estimate the electric field developed by considering a spherical grain of
mean size $a =0.5\mu$m carrying $q=3\cdot10^3\,e$ charges (Desch \&
Cuzzi 2000). This gives
\begin{equation}
E = \frac{q}{4\pi \varepsilon_0 a^2} = 20 {\rm MV m^{-1}}
\end{equation}
where $\varepsilon_0$ is the
electrostatic field constant, and $e$
is the elementary charge.
Dowds, Barrett \& Diver (2003) demonstrate the
effectiveness of a succession of these avalanches (known as a {\it streamer} in ionising
the nitrogen gas between two capacitor plates for a field strength of
only 5MV\,m$^{-1}$.  Li et al.~(2007) and Li, Ebert \& Brok (2008)
support this finding for an electric field strength of 10 MV\,m$^{-1}$
from their study 
based on different numerical methods. Hence, the field strength
between charged grains is sufficient to produce streamers that enhance the
local degree of gas ionisation inside the dust cloud. The electric
field will be stronger if the grain's shape depart from spherical
symmetry (Stark, Potts \& Diver 2006).  

 The streamers that form are a growing ionisation front that
 propagates into non-ionised matter.  They are present in lightning
 ladders and they can emerge from ionisation avalanches in 
 weakly ionized plasmas in free space. Such field breakdowns are
 stochastic and non-linear processes that produce an exponentially
 increasing number of charges. Streamers are self-sustaining
 (hence an instability), and each streamer is ignited by the large
 number ($\sim 10^6$) of free charges produced by the avalanche of the
 accelerated electron from one gas-phase electron-ion pair. Between
 $10^{13}$ and $10^{14}$ cm$^{-3}$ free charges form during the
 consecutive avalanches in the streamer (Dowds, Barrett \& Diver 2003,
 Li et al. 2007). One streamer alone is therefore capable of
 satisfying $I_{\rm dust}\gg R_{\rm dust}$
locally, and hence
causing departure from local charge equilibrium. 

Such local ionisation events are only likely to be capable of
explaining the observations of Brown Dwarf flares and radio emission
if they occur over a sufficiently large volume of the atmosphere. The
cloud grains, however, are confined to a certain volume of the
atmosphere by thermodynamic constrains. This volume increases with
decreasing surface gravity as we have shown in
Sect.~\ref{sec:cloudheight}. The cloud height would be the maximum
size that a streamer leading to a lightning bolt could achieve in the
case of intra-cloud lightning. In this volume, the simultaneous
occurrence of discharge in streamers will occur, and the gas in the
cloud could then locally couple to the magnetic field for a short time
appearing as e.g. stochastic flares.  Over a recombination time, we
expect that these excess charges will recombine onto dust grains.

\subsection{Neutralisation  versus avalanche time scales}\label{ss:neutralis}

We have argued that dust grains populate Brown Dwarf and extrasolar
 planet atmosphere over a extension that varies mainly with
 gravity. We have further argued that non-linear processes are needed
 to produce discharge processes such as those in solar system planets,
 and we have suggested exponential electron avalanche as an attractive
 possibility.  It remains to be shown if the grains could be
 neutralised by gas-dust collisions quickly enough to prohibit the
 built-up of the streamer-avalanche mechanism that leads to the
 appearance of lightning bolts.

\begin{figure}
\begin{center}
\resizebox{16.5cm}{!}{\includegraphics{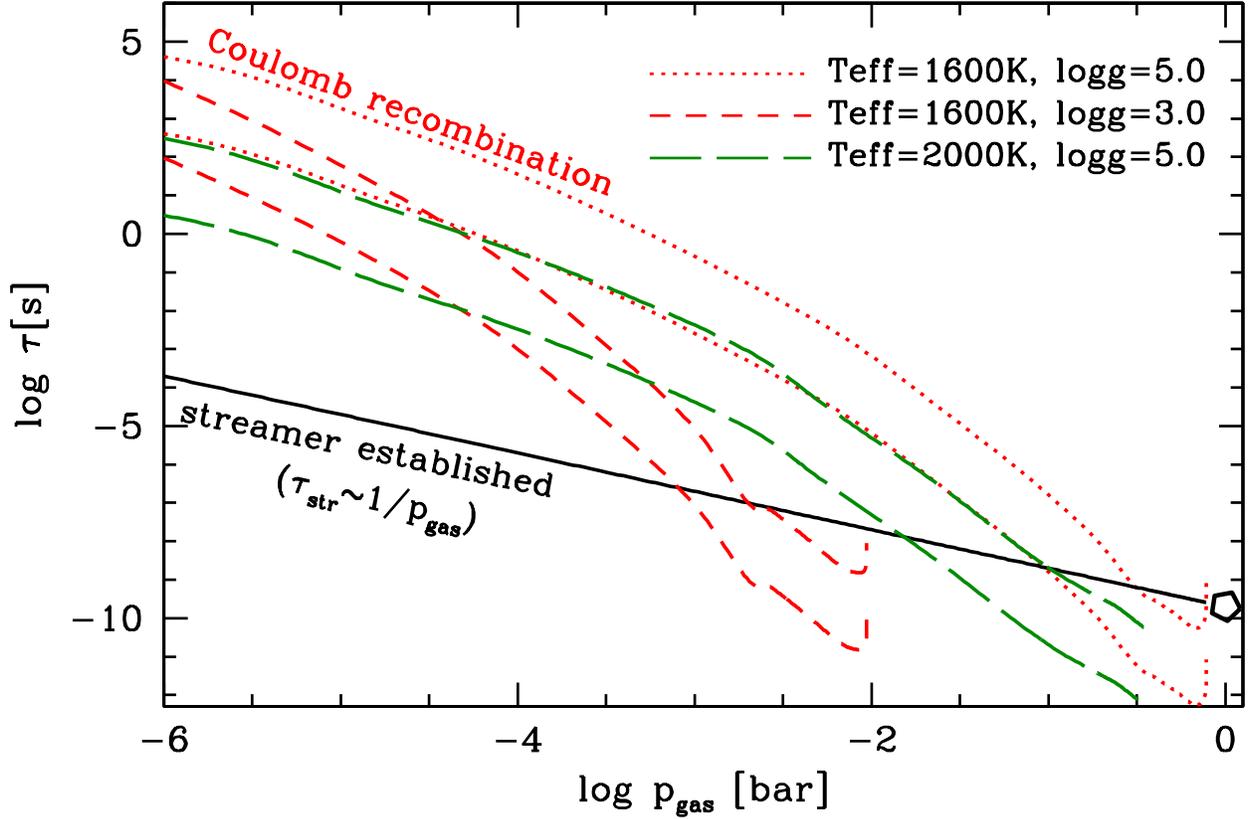}}
 \caption{\small The time scale of grain neutralisation by freely
 impinging electrons $\tau_{\rm recom}^{\rm cgs}$ (red and green
 dashed and dotted lines) compared to the time required to establish a
 streamer $\tau_{\rm str}$ (black solid line). The neutralisation time
 scale $\tau_{\rm str}$ is shown for three different {\sc
 Drift-Phoenix} atmosphere models (hot/cold and high/low log(g)) which
 are each evaluated for two cases: charge load of small grains
 ($q=10\,e$, upper line) and larger grains ($q=10^2\,e$, lower line).
 The black symbol is the numerical data from Dowds, Barrett \& Diver
 (2003) extrapolated such that $\tau_{\rm str}\sim 1/p_{\rm gas}$ (Eq.~\ref{eq:taustr}).  }
\label{fig:taurc}
\end{center}
\end{figure}

Dowds, Barrett \& Diver (2003) carried out numerical experiments where
the electron avalanche process and the development of streamers were
studied between two 0.1cm-size capacitor plates with a plate distance
of $5\mu$m. The capacitor was filled with nitrogen gas of a typical
terrestrial atmospheric pressure of 1bar. They show that a streamer is
fully developed after about $\tau_{\rm str}=$0.2ns ($=2\cdot
10^{-10}$s; green symbol, Fig.~\ref{fig:taurc}), a time scale
confirmed by comparable works of Li, Ebert \& Brok (2008) and by 3D
simulations of Li, Ebert \& Hundsdorfer (2009). Following Ebert et
al. (2010), we scale $\tau_{\rm str}$ with the local gas pressure as
$\tau_{\rm str}\sim 1 /p_{\rm gas}$ (solid black line,
Fig.~\ref{fig:taurc}, \footnote{The time scale to
establish a streamer can also be defined as
\begin{equation}
\label{eq:taustr}
\tau_{\rm str} = \frac{N(x(t))}{ \frac{dN(x(t))}{dt}},
\end{equation}
with $N(x(t)) = N_0\exp(\frac{x(t)}{\lambda_{\rm e}})$ the exponential
growth of charges in a streamer (e.g. Braithwaite 2000). With $N_0$
the initial values of number of charges dropping out of
Eq.~\ref{eq:taustr}, by assuming a stationary electron velocity
($dx(t)/dt=const$) through the capacitor, and with the electron's mean
free path $\lambda_{\rm e} \sim 1/p_{\rm gas}$ follows that $\tau_{\rm
str}\sim 1 /p_{\rm gas}$.}). 

To estimate the grain charge neutralisation time scale we adopt the
neutralisation rate given by Desch \& Cuzzi (2000; Eq. 24) which
assumes that electrons can freely impinge on the positively charged
grain's surface if the Coulomb energy of the grain is larger than the
thermal kinetic energy of the electrons. This time scale provides an
estimate for efficient neutralisation as electrons have a higher mobility
than the much heavier ions. From their formula, one derives the
neutralisation time scale for a grain of size $a$ (Eq. 25 in Desch \&
Cuzzi 2000) in cgs units\footnote{The neutralisation time scale is in
SI units ($e=1.602\cdot10^{-19}$C),
\begin{equation}
\label{eq:trecombSI}
\tau_{\rm recom}^{\rm SI}= \frac{4\pi\epsilon_0}{q^2}\frac{k T}{n_{\rm e}c_{\rm e} \pi <a>}. 
\end{equation}
} applying
the mean grain size, $<a>$, from our {\sc Drift-Phoenix} model
\begin{equation}
\label{eq:trecombcgs}
\tau_{\rm recom}^{\rm cgs}= \frac{1}{q^2}\frac{k T}{n_{\rm e}c_{\rm e}\pi <a>},
\end{equation}
where $n_{\rm e}$ is the gas-phase electron density, $c_{\rm
e}=\sqrt{2\,k\,T/m_{\rm e}}$ is the electron thermal velocity and
$m_{\rm e}$ is the electron mass.  We show two limiting cases for the
grain charge $q$ according to Fortov et al. (2001) where small grains
carry of the order of $q=10\,e$ (each uper line) and large grains
$q=10^2\,e$ (each lower line).  The elementary charge is in units of
$e=4.803\cdot10^{-10}$statC. We also include the recombination
time-scales for a hotter (T$_{\rm eff}=2000$K, log(g)=5.0; green long
dashed) and a lower-gravity atmosphere model (T$_{\rm eff}=1600$K, log(g)=3.0;
red short dashed) for comparison. The time scale for dust grain neutralisation by
impinging electrons is longer than the time needed to establish a
streamer (solid black line) throughout a significant fraction of the
cloud in all the model atmospheres considered
(Fig.~\ref{fig:taurc}). Only the largest particles at the cloud base
would neutralise too quickly for the streamer mechanisms to kick in
efficiently if the neutralisation process proceeds without collisions
as assumed in Eq.~\ref{eq:trecombcgs}. We note that this is also the
part of the cloud where the largest particles accumulate and quickly
evaporate inside a narrow cloud layer. These grains are almost pure metal
as they are made of Fe[s] with small inclusions of TiO$_2$[s] and
Al$_2$O$_3$[s] followed by a thin cloud region with almost homogeneous
Al$_2$O$_3$[s] grains ([s] -- solid).


\section{Conclusions}

Brown Dwarf and extrasolar planet atmospheres form clouds which
strongly influence the local chemistry and physics. These clouds need
to form from the gas-phase as Brown Dwarfs and most of the planets do
not have a crust from which seed particles can be swept up. These
clouds are globally neutral obeying dust-gas charge equilibrium. On
short time scales, however, stochastic ionisation events may occur as observed
in the solar system planets.  We have argued that a large
part of the clouds in Brown Dwarfs and extrasolar planets is
susceptible to local discharge events which are triggered by charged
dust grains. Such discharges occur on time scales shorter than the
time required to neutralise the dust grains, and their superposition
might produce enough free charges to suggest a partial and stochastic
coupling of the atmosphere to a large-scale magnetic field. Discharge
processes in Brown Dwarf and exoplanetary atmospheres can not connect
to a crust as on terrestrial planets, hence, they will experience
intra-cloud discharges comparable to volcano plumes and dust devils.

\bigskip
We thank the referee for critical comments and useful suggestions. ChH
thanks the KITP Exoplanet program (NSF grant PHY05-51164) where part
of this work was performed.


\label{lastpage}

\clearpage



\clearpage

\clearpage

\begin{table}
\begin{center}
\caption{\small Cloud extension, $\Delta H_{\rm cloud}$, for Brown Dwarfs and
WASP-like giant gas planets ($R_{\rm J}=7.1492\cdot\,10^9$cm --
equatorial Jupiter radius; R$_{\odot}=6.9598\cdot 10^{10}$cm -- solar
radius) as derived  from the {\sc Drift-Phoenix} atmosphere simulations results in Figs.~\ref{fig:cloudthickness}.  All T$_{\rm eff}$ for the WASP planets are taken as A=0 (A - albedo; see
references in table).} \label{tab:examples}
\begin{tabular}{c c|c c}
\multicolumn{2}{c}{\bf WASP 12b}  & \multicolumn{2}{c}{\bf WASP 15b}\\[-0.1cm]
\multicolumn{2}{c}{(Hebb et al. 2009)} &\multicolumn{2}{c}{(West et al. 2009)}\\
\hline
T$_{\rm eff}$=2516K       &  \hspace*{-0.5cm}{\bf cloud extension:} & T$_{\rm eff}$= 1652K &  \hspace*{-0.5cm}{\bf cloud extension:} \\
$\log$(g) =  2.9           &  0.03  R$_* =$  & $\log$(g) = 2.8        &  0.04  R$_* =$\\
R$_*$  = 1.79 R$_{\rm J}$ &  {\bf 3820km}      & R$_*$  = 1.43 R$_{\rm J}$&  {\bf 3830km}\\
                     &                    &   &    \\
\multicolumn{2}{c}{\bf WASP 10b}  & \multicolumn{2}{c}{\bf WASP 14b}\\[-0.1cm]
\multicolumn{2}{c}{(Christian et al. 2009)} &\multicolumn{2}{c}{(Joshi et al. 2009))}\\
\hline
T$_{\rm eff}$=1119K       &  \hspace*{-0.5cm}{\bf cloud extension:} & T$_{\rm eff}$= 1866K &  \hspace*{-0.5cm}{\bf cloud extension:}\\
$\log$(g) =  3.6           & 0.011  R$_* =$      & $\log$(g) = 4.0        & 0.005   R$_* =$\\
R$_*$ = 1.28R$_{\rm J}$   & {\bf 1000km}              & R$_*$  = 1.28R$_{\rm J}$ & {\bf 410km}\\
                     &                    &   &    \\
\multicolumn{4}{c}{\bf binary system 2MASS J05352184-0546085}\\[-0.1cm]
\multicolumn{4}{c}{(Stassun et al. 2007)}\\
\hline
T$_{\rm eff,1}\sim$ 2700K &  \hspace*{-0.5cm}{\bf cloud extension:} & T$_{\rm eff,2}\sim$2800K   &  \hspace*{-0.5cm}{\bf cloud extension:}\\
$\log$(g)$_1$ = 3.5             & 0.012  R$_* =$  & $\log$(g)$_2$ = 3.6          & 0.011  R$_* =$\\
R$_{*, 1}$ = 0.67R$_{\odot}$      & {\bf 5620km}          & R$_{*, 2}$ = 0.48R$_{\odot}$ & {\bf 3710km}\\
\end{tabular}
\end{center}
\end{table}







\clearpage

\end{document}